\documentclass[pre,showpacs,preprintnumbers,amsmath,amssymb,twocolumn]{revtex4}
\usepackage{graphicx}

\usepackage{amsmath}
\usepackage{amssymb}
\usepackage{amsthm}
\usepackage{mathrsfs}

\begin{document}

\title{Current amplification and magnetic reconnection at a 3D null point. I -
Physical characteristics}

\author{D. I. Pontin}
\email{dpontin@maths.dundee.ac.uk}
\affiliation{ 
Space Science Center, University of New Hampshire, USA 
\\(now at Department of Mathematics, University of Dundee, Dundee, Scotland)
}%

\author{K. Galsgaard}%
\affiliation{%
Niels Bohr Institute, University of Copenhagen, Copenhagen, Denmark }%



\begin{abstract}
  The behaviour of magnetic perturbations of an initially potential
  three-dimensional equilibrium magnetic null point are investigated. The
  basic components which constitute a typical disturbance are taken to be
  rotations and shears, in line with previous work. The spine and fan of the
  null point (the field lines which asymptotically approach or recede from the
  null) are subjected to such rotational and shear perturbations, using
  three-dimensional magnetohydrodynamic simulations. It is found
  that rotations of the fan plane and about the spine lead to current sheets
  which are spatially diffuse in at least one direction, and form in the
  locations of the spine and fan. However, shearing perturbations lead to
  3D-localised current sheets focused at the null point itself. In addition,
  rotations are associated with a growth of current parallel to the spine,
  driving rotational flows and a type of rotational reconnection. Shears,
  on the other hand, cause a current through the null which is parallel to the 
  fan plane, and are associated with stagnation-type flows and field line
  reconnection across both the spine and fan. The importance of the parallel
  electric field, and its meaning as a reconnection rate, are discussed.
\end{abstract}


\maketitle

\section{Introduction}
Many important physical phenomena in astrophysical plasmas are powered by
magnetic reconnection. The locations at which reconnection is likely to occur
in a complex three-dimensional magnetic field are those regions where strong
currents (possibly singular in the ideal regime) may develop. However,
determining what these locations might be is a non-trivial problem.

There are a growing number of theories and pieces of evidence to suggest that
3D null points and separators (magnetic field lines which join two such nulls)
may be sites of preferential current growth, in both the Solar corona and the
Earth's magnetosphere. The field topology in the vicinity of such a 3D null
point is defined by the field lines which asymptotically approach (or recede
from) the null. These fall into two categories. A single pair of field lines
approach (recede from) the null from opposite directions, and are termed the
`spine' (or $\gamma$-line), while an infinite family of field lines recede
from (approach) the null in a surface called the fan (or $\Sigma$-)
plane. 
The spine and fan of a given null may be determined by
examining the linear field topology near the null, defined by the equation
\begin{displaymath}
{\bf B} = {\cal M} \cdot {\bf r},
\end{displaymath}
where the matrix ${\cal M}$ is given by  the Jacobian of ${\bf B}$
\citep[e.g.][]{fukao1975}. The eigenvectors of ${\cal M}$ (whose corresponding
eigenvalues sum to zero since $\nabla \cdot {\bf B}=0$) now define the spine
and fan. The two eigenvectors whose eigenvalues are of like sign (or whose
real parts have like sign) lie in the fan plane, while the third points along
the spine.  The fan surface is a separatrix surface of the magnetic field,
separating unique topological regions.  While the spine does not separate
topological regions (being only a line), it is nonetheless an important
geometrical feature of the field. This is because, firstly, magnetic field
lines converge on (or diverge from) it, and secondly, it is a field line along
which disturbances may be channelled towards the null (as we shall see later).

3D null points are predicted to be present in abundance in the solar corona.
It should however be noted that many recent models of the magnetic field above
  the solar surface are based upon the Magnetic Charge Topology approach,
  extrapolating a potential field from point magnetic sources in the
  photosphere. This is clearly a great simplification, and many (or all) of the
  `photospheric null points' in such models will not be present when
  these idealised flux sources are replaced by more realistic finite flux
  patches. However, 
  various approaches also predict the presence of 3D null points up in the corona,
  with an average of between approximately 7 and 15 coronal nulls
  expected for every 100 photospheric flux concentrations
  \citep[][]{schrijver2002,longcope2003,close2004}. The presence of these
  coronal nulls is expected to be more robust to the method of field
  extrapolation \citep[][]{brown2000,brown2001}.
The separatrix surfaces and
separators associated with these coronal null points are thought to be likely
sites of coronal heating and reconnection 
\citep{longcope1996, priest2005}. Furthermore, there is observational evidence
that reconnection at a 3D null point (both fan-type and spine-type
reconnection) may be at work in  some solar flares \citep{fletcher2001}.
Closer to home, separator reconnection is thought to occur on the dayside of
the Earth's magnetosphere \citep[e.g.][]{siscoe1988}. In addition, there has
been a recent in situ observation by the Cluster spacecraft \citep{xiao2006}
of a 3D magnetic null, at which it is proposed that reconnection is occurring,
located within the current sheet in the Earth's magnetotail.

Although it is now realised that 3D nulls are of great importance for
reconnection in realistic 3D geometries, what is still lacking is a clear
picture of what reconnection processes at such nulls look like. In particular,
a description of the sorts of physical signatures expected (current sheets,
plasma flows etc.), in order to lead the analysis of new observations, is of
great importance. Our aim in this paper is to go some way to providing such a
picture.

While the relationship between reconnection at a separator and a single 3D
null point is not well known, it is clear that the two should be linked in
some way. What is known is that both are prone to collapse in response to
external motions
\citep[e.g.][]{longcopecowley1996,bulanov1984,pontincraig2005}. That is, a 3D
null point is a `weak point' of the magnetic field to which disturbances are
attracted \citep{galsgaard1997,galsgaardreddy1997}.  Furthermore, kinematic
considerations suggest that null points and separators are both locations
where singularities may form in ideal MHD \citep[e.g.][]{lau1990,priest1996}.
The attraction of disturbances to 2D nulls is also well documented
\citep[e.g.][]{hassam1992,craig1993}.

The local behaviour of a perturbed 3D magnetic null point has been examined in
the linear regime for the cold, resistive MHD equations by
\cite{rickard1996}. They employed a modal decomposition, and determined that
only $m=0$ and $m=1$ modes can lead to currents at the null point, where $m$
is the azimuthal wavenumber. In addition, they found that while $m=0$ modes
are attracted to the spine and fan of the null, $m=1$ type perturbations tend
to focus in towards the null itself. This was demonstrated in the linear
regime with 2D simulations in the $rz$-plane. One of the major aims of the
present work is to ascertain whether such behaviour for the evolution of
disturbances is found in the full MHD regime, using 3D simulations.

In addition to the above, we will investigate which current components
develop at the null in response to different perturbations, and the
implications which this has for the plasma flow and field line behaviour. 
In each case we consider, the plasma is initially at rest, and so with no
  flow through our 
  boundaries, we do not compare with strongly driven flux pile-up models
  \citep[e.g.][]{craig1996}. 
It
is anticipated that different current orientations at the null will lead to very
different behaviour \citep{pontin2004, pontinhornig2005}. One of the present
simulations is very similar to that described in the work of
\cite{galsgaardpriest2003}, and so is only briefly summarised in what
follows. The other simulations combine with this to create a complete picture
of the subject at hand. 

In Section \ref{setup} we describe the numerical scheme and the initial
conditions used. In Sections \ref{fanrotsec}--\ref{spineshsec} we describe the
results of our simulations for various different types of perturbations, and
in Section \ref{summary} we give a summary and conclusions.

\section{Numerical scheme and simulation setup}\label{setup}
\subsection{Numerical Scheme}
The numerical scheme employed in the simulations which follow is described
briefly below (further details may be found in \cite{nordlund1997} and at
http://www.astro.ku.dk/$\sim$kg). We solve the three-dimensional resistive MHD
equations in the form
\begin{eqnarray}
\frac{\partial {\bf B}}{\partial t} & = & - \nabla \times {\bf E},
\label{numeq1}\\ 
{\bf E} & = & -\left( {\bf v} \times {\bf B} \right)
\: + \: \eta {\bf J}, \label{numeq2}\\ 
{\bf J} & = & \nabla \times
{\bf B}, \label{numeq3}\\ 
\frac{\partial \rho}{\partial t} & = & -
\nabla \cdot \left( \rho {\bf v} \right), \label{numeq4}\\
\frac{\partial}{\partial t}\left( \rho {\bf v} \right) & = & - \nabla
\cdot \left( \rho {\bf v} {\bf v} \: + \: {\underline {\underline
\tau}} \right) \: - \: \nabla P \: + \: {\bf J} \times {\bf B},
\label{numeq5}\\ 
\frac{\partial e}{\partial t} & = & -\nabla \cdot
\left( e {\bf v} \right) \: - \: P \: \nabla \cdot {\bf v} \: + \:
Q_{visc} \: + \: Q_{J} \label{numeq6},
\end{eqnarray}
where ${\bf B}$ is the magnetic field, ${\bf E}$ the electric field,
${\bf v}$ the plasma velocity, $\eta$ the resistivity, ${\bf J}$ the
electric current, ${\rho}$ the density, ${\underline {\underline
\tau}}$ the viscous stress tensor, $P$ the pressure, $e$ the internal
energy, $Q_{visc}$ the viscous dissipation and $Q_{J}$ the Joule
dissipation. An ideal gas is assumed, and hence $P \: = \: \left(
\gamma -1 \right) \: e \: = \: {\textstyle \frac{2}{3}}e$.

The equations above have been non-dimensionalised by setting
the magnetic permeability $\mu_0 = 1$, and the gas constant ($\mathcal{R}$)
equal to
the mean molecular weight ($M$). The result is that, for a cubic domain of
unit size, if $| \rho |=| {\bf B} | = 1$, then time is measured in units of the
Alfv\'{e}n travel time across the domain ($\tau_A=L \sqrt{\mu \rho_0}
/ B_0$, where $L$ is the size of the domain, and $\rho_0$ and $B_0$
are typical values of the density and magnetic field respectively). 

The equations are solved on staggered meshes; in this way the required MHD
conservation laws are automatically satisfied. Spatial derivatives are
evaluated using a sixth-order-accurate finite difference method. It is often
the case that, due to the staggered mesh, this value of the derivative is
returned in exactly the position it is needed. When this is not the case, the
values are calculated using a fifth-order interpolation method at the relevant
position.  A non-uniform mesh is used in the simulations, with higher
resolution in the centre of the domain in each coordinate direction. In this
way it is possible to better resolve behaviour in the vicinity of the null,
and indeed across the spine and fan, while keeping the boundaries `far away'
to reduce their effect.  A third-order predictor-corrector method is employed
for time-stepping. All simulations are carried out on grids of $128^3$ or
$256^3$ resolution.

Viscosity is handled using a combined second-order (constant $\nu$)
and fourth-order method, which is capable of providing sufficient localised
dissipation where necessary to handle the development of numerical
instabilities. The result is that it is possible to achieve much lower
effective values of $\nu$ for a given numerical resolution than
with a constant-$\nu$ approach. This is achieved by having an enhanced viscosity
at length scales close to the numerical resolution limit, effectively dissipating
short wavelength disturbances, while leaving larger-scale structures nearly
undamped. Viscosity defined in such a way is often termed
`hyper-viscosity'. Such an approach is 
used in 3D simulations to maximise the  fraction of the 3D domain that has an
`ideal' behaviour. Two different models are used for the resistivity; either
a traditional constant-$\eta$ (second-order) model or a hyper-resistivity model 
(labelled $\eta_h$ in the text) similar in approach as the viscosity model. 
Comparisons between identical simulation runs using the two resistivity
approaches show the same general evolution, with the main difference being
that the hyper-resistivity runs show more spacially localised structures.
This is demonstrated explicity for one particular case below (see Section
  \ref{fanshsec}). 

In a number of the simulation runs, `trace particles'  are used in order to
track the motion of field lines in time. These points are chosen at $t=0$, and
are then followed in time throughout the simulation by integrating the plasma
flow field. These `fluid elements' are then used to define field line
footpoints.

Finally, the boundary conditions are closed in all three directions.  In
addition, as we aim to study only the initial localisation of perturbations at
the null, a boundary damping zone is included, in order to limit the
reflection of waves back into the domain from the boundary.  Within this
region, a fraction of the kinetic energy is removed per unit time, which, when
chosen appropriately for the wave speed, may effectively damp a large portion
of the wave energy.
The effect of the boundaries on the dynamics of the null point is, on the other 
hand, negligible, as the simulations are all terminated before the perturbations 
have time to reflect off them and reach back to the region around the null again.

\subsection{Initial setup}
The simulation is set up as follows. The initial `background' magnetic field is
potential, and describes a 3D magnetic null point located at the origin,
specifically ${\bf B}=B_0 \left( x, y, -2z \right)$. The plasma is initially at
rest, with $\rho=1$, $e=5\beta/2$ everywhere, 
where $\beta$ determines the plasma-$\beta$. 
The system is then disturbed by perturbing
the magnetic field. We choose to add a magnetic field instead of, say,
imposing some velocity on a line-tied boundary, so as to examine the evolution
of the null point, and the reconnective behaviour, in an undriven
situation. We focus on the behaviour of the transient pulse as it moves
towards the null. Note that the fact that there is no flow through our
  boundaries means that we do not compare with strongly driven flux pile-up
  models. 

In order
for the null point to be affected, the perturbation must disturb either the
spine or fan of the null, otherwise the disturbance will simply propagate 
back and forth along the associated magnetic field lines, bouncing between the 
boundaries. The perturbation magnetic
field must of course be divergence-free, and in each of the cases discussed
takes the general form
\begin{equation}\label{perturb}
\begin{array}{rcl}
b_i & = & - b_0 R_1 \; {\rm sin(\theta_1)} \; exp\left( 
  - \frac{\left(R_1-R_{10}\right)^2}{{a_h}^2} 
- \frac{\left(\zeta-\zeta_0\right)^2}{{b_h}^2} \right)\\
b_j & = & b_0 R_1 \; {\rm cos(\theta_1)} \; exp\left( 
  - \frac{\left(R_1-R_{10}\right)^2}{{a_h}^2} 
- \frac{\left(\zeta-\zeta_0\right)^2}{{b_h}^2} \right),
\end{array}
\end{equation}
where $b_i$ and $b_j$ are the two components of the perturbation magnetic
field, $\zeta$ is the third spatial coordinate, and $b_0$, $R_{10}$, $\zeta_0$, $a_h$
and $b_h$ are constants. $R_1$ and $\theta_1$ are
defined depending on the orientation of the perturbation. The
perturbation is localised by the exponential
terms within either a linear tube or a torus.

\begin{figure}
\centering
\noindent\includegraphics[width=20pc]{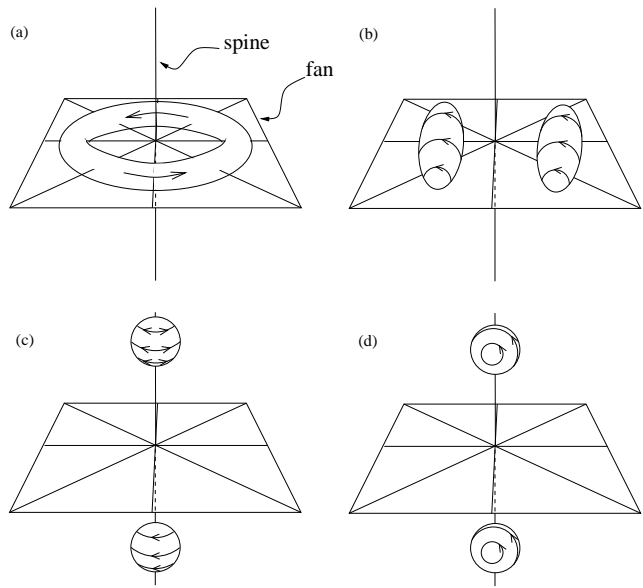}
\caption{Schematic view of isosurfaces of the different perturbation magnetic
  fields, with arrows indicating the field orientation. (a)
rotation within the fan plane, (b) shear in the fan plane (c) rotation of the
spine, in the same or opposite sense above and below the fan, and (d) shear of
the spine. In each case the vertical line is the (undisturbed) spine of the
null, and the square is the fan.}
\label{perturb_pic}
\end{figure}

\begin{figure}
\centering
\noindent\includegraphics[width=20pc]{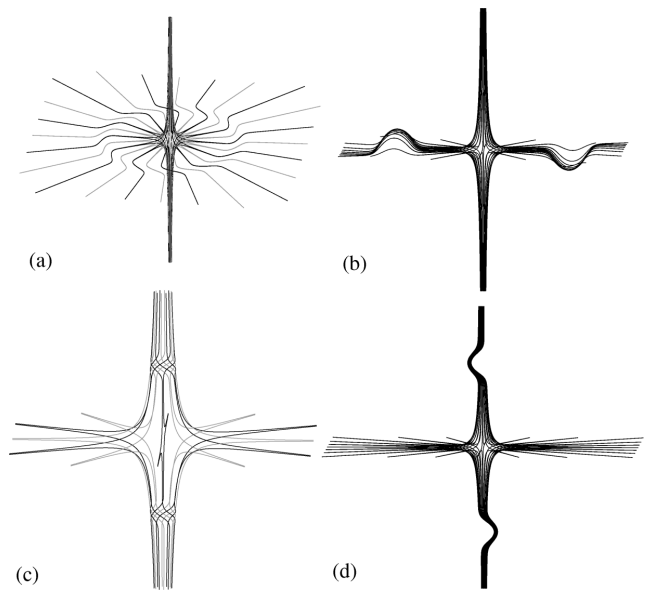}
\caption{Sample magnetic field lines at $t=0$ for the different
  perturbations, (a)-(d) as in Figure \ref{perturb_pic}.}
\label{perturb_bl}
\end{figure}

Five different types of perturbation will be considered. These are described
briefly below, and are illustrated in Figures
\ref{perturb_pic} and \ref{perturb_bl}. Figure \ref{perturb_pic} shows isosurfaces of the
disturbance field magnitude, with the arrows showing the direction of the
field. Figure \ref{perturb_bl} plots a selection of field lines from around the null, showing
the effect of the disturbances on the the magnetic field. The first type of
perturbation corresponds to the $m=0$ 
disturbances of \cite{rickard1996}, that is, rotational motions. We consider
two cases, in the first of which (Section \ref{fanrotsec}) the rotation is
concentrated in the fan plane (but away from the null, see Figures
\ref{perturb_pic}(a), \ref{perturb_bl}(a)). One may also perform a rotation
about the spine, but 
away from the fan plane, either in the same or opposite sense above and
below the fan (see Figures \ref{perturb_pic}(c), \ref{perturb_bl}(c)). This
has been investigated in 
detail by \cite{galsgaardpriest2003}, and is discussed briefly in Section
\ref{spinerotsec}. Alternatively, the null point separatrices may be perturbed
by applying some shear (corresponding to $m=1$), to either the fan (Section
\ref{fanshsec}, Figures \ref{perturb_pic}(b), \ref{perturb_bl}(b)) or the
spine (Section 
\ref{spineshsec}, Figures \ref{perturb_pic}(d), \ref{perturb_bl}(d)). Any more
generic perturbation 
may be made up from a combination of such rotations and shears.

\section{Rotation in the fan plane}\label{fanrotsec}
For a rotation in the fan plane, the perturbation magnetic field lies in the 
$xy$-plane, such that $\left\{b_i, b_j\right\} = \left\{b_x,b_y
\right\}$ and $\zeta=z$ in Equation (\ref{perturb}), and in addition we take 
$R_1=\sqrt{x^2+y^2}$, $\theta_1={\rm tan}^{-1}(y/x)$, and
$\zeta_0=0$. Thus the current is initially concentrated in a torus whose
toroidal axis lies in the fan plane at a radius $R_{10}$ from the origin
(null), see Figures \ref{perturb_pic}(a) and \ref{perturb_bl}(a). 
The domain size is chosen to be  $1.5\times1.5\times2.5$, in
order to limit the effect of the upper/lower boundaries in $z$.
We use the hyper-resistive model, $\eta=\eta_h$, and the characteristic
parameters for the experiment are;
$B_0=1, b_0=0.1, R_{10}=0.16, \beta=0.01, a_h=b_h=0.06$, giving
a travel time to the top ($z$) boundary of order 1.8 and to the $xy$-boundaries of
1.5 in code units, for the main body of the disturbance.

\subsection{Current density evolution}

\begin{figure}
\centering
\noindent\includegraphics[width=20pc]{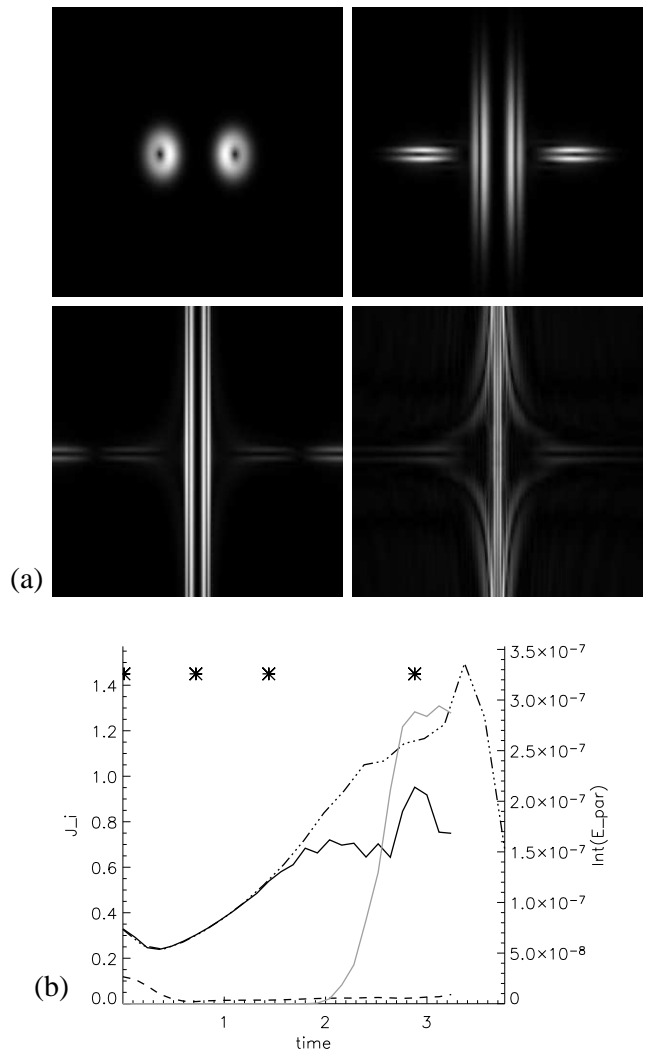}
\caption{
  For a rotation of the fan; (a) current modulus in the $y=0$
  plane for $[x,z]=[\pm0.75,\pm0.75]$, at 
  times marked by asterisks in (b) ($t=0, 0.72, 1.44, 2.88$).
  (b) Evolution of the maximum value of each current component
  ($J_x$ and $J_y$ dashed, $J_z$ solid, black) and of the integrated 
  parallel electric field along the spine (grey). Domain size is
  $1.5\times1.5\times2.5$  and parameters used are
  $B_0=1, b_0=0.1, R_{10}=0.16, \beta=0.01, a_h=0.06, b_h=0.06$ and $\eta=\eta_h$.
  The dot-dashed line shows $J_z$ evolution in
  a run with the same parameters, but at $256^3$ resolution.}
\label{j_fanrot}
\end{figure}

Figure \ref{j_fanrot} shows the time evolution of the current modulus in a
plane through the null and spine axis (chosen to be the $y=0$ plane, although
we have cylindrical symmetry away from the boundaries). It is evident from the
image that pulses propagating both inwards and outwards develop, as one would expect.
In the following, we refer to the disturbances which propagate towards and
  away from the null point as the `ingoing' and `outgoing' pulses,
 respectively. 
Note that here, as in the simulations which follow, both the ingoing pulse
and outgoing  pulse have a somewhat complicated structure. Specifically, it
appears from Figure \ref{j_fanrot}(a) that there are two wavefronts which
localise towards the null, 
although in fact these two wavefronts (in $|{\bf J}|$) correspond to current
concentrations of 
opposite sign (in $J_z$), which demark strong field gradients at the `back
end' and `front end' of the single ingoing `pulse'.  The magnetic field
gradients must necessarily pass through zero at the centre of this pulse,
hence the appearance of two strong bands in $|{\bf J}|$.
The ingoing pulse propagates along the
`background' field lines, and concentrates in a current tube centred on the
spine. There is, however, no preferential attraction of the current to the null
itself. It is interesting that even though the field structure is
hyperbolic, the current localises very uniformly in $z$, i.e.~the wavefront is
very close to vertical, as would be expected for an Alfv{\' e}n wave.
Note finally that in the last image there are some
unavoidable reflections of the disturbance from the upper boundary back along
the background field lines. We have checked that the boundaries play no significant
role in the qualitative or quatitative evolution by repeating the simulation
in a domain of dimensions $1.5\times1.5\times4.5$ (which increases the
perturbation travel time to and from the $z$-boundaries to greater than 4 in
code units).

Further insight may be gained by examining the evolution of the different
components of the current. The maximum value of each component is plotted in
Figure \ref{j_fanrot}(b) (we take the maximum within the region
$[\pm0.25,\pm0.25,\pm0.42]$ in each direction to exclude boundary effects).
The figure shows that as the current localises towards the spine, it is $J_z$,
i.e.~the current parallel to the spine, which is significantly amplified,
while $J_x$ and $J_y$ are not. This is to be expected, as it is $dx$ and $dy$
which are decreasing as the disturbance is squeezed in towards the spine.
The current eventually reaches a maximum as it localises.  
This is due to a combination of the finite numerical resolution
and the imposed resistivity model. For constant $\eta$, increasing the numerical
resolution ($N$) will eventually result in a behaviour of the solution that is 
independent of $N$. On the other hand, the current in the hyper-resistive 
case will continue to increase with $N$, with the physical structure having 
a length scale comparable to the numerical resolution. 
This can be seen by
examining the growth of the dominant ($z$) component of ${\bf J}$ for a run
with double the resolution, shown by the dot-dashed line in Figure
\ref{j_fanrot}(b). The early evolution is very similar, though a higher
current peak is eventually achieved after a longer period of localisation.

In a truly ideal evolution, it is expected that the current would
increase indefinitely in time, although more and more of the energy
associated with the disturbance would `escape' down the spine
\citep[c.f.][]{mclaughlin2004}.  However, in a true physical case, no matter
how small the resistivity, 
it will always become important eventually, once the current becomes
sufficiently intense, and as a result some energy can be dissipated. This will
occur all along the spine.

One very interesting question is how the maximum current depends on the
resistivity. It is impossible here for us to use realistic resistive
parameters appropriate for, e.g.~the Solar corona, where the resistivity is
typically of the order $10^{-14}$. One crucial feature of any reconnection
model is therefore whether the reconnection rate scales as some negative power
of $\eta$---if so then it will typically correspond to `slow' reconnective
behaviour in a realistic plasma. It is our intention to investigate this
dependence in a subsequent paper in this series. It has furthermore recently
been demonstrated \citep{craig2005,pontincraig2005} that the plasma pressure
(i.e.~the value of the plasma-$\beta$) can have a profound effect  on current
scalings at 2D and 3D null points. This will also be investigated further in a
later paper.

At a first glance, the type of tubular current structure described above is
reminiscent of those found in the incompressible `spine reconnection'
solutions first described by \cite{craig1996} (in their case there exist two
tubes in close proximity due to the assumed symmetry). In a general
incompressible time-dependent case, a single tube of current is stretched out
along the spine of the null, with the current being purely azimuthal within
the tube (that is, directed parallel to the fan plane),  rather than axial
\citep{pontincraig2006}. This is clearly rather different from the  situation
we have here, where the current is very close to being parallel to the
spine. In addition, the incompressible spine reconnection solutions are
associated with reconnection of field lines across the spine and fan
(advection across the fan and diffusion across the spine), which we shall see below
does not occur here.

One final important characteristic of the current evolution observed in our
simulation is a non-linear coupling which may occur when the perturbation
magnetic field is strong compared with the background field. As observed by
\cite{galsgaardpriest2003}, the main Alfv{\' e}nic disturbance may also couple
to a fast-mode wave which is attracted to the null point itself. This is,
however, a fairly weak effect. \cite{galsgaardpriest2003} found that the
coupling only existed in cases where the boundary driving was ramped sharply
up from zero, and similarly we find that the strength of (and gradients in)
the disturbance field must be very strong in order that the effect can be seen
at all.

\subsection{Plasma flow and field line behaviour}
\begin{figure}
\centering
\noindent\includegraphics[width=20pc]{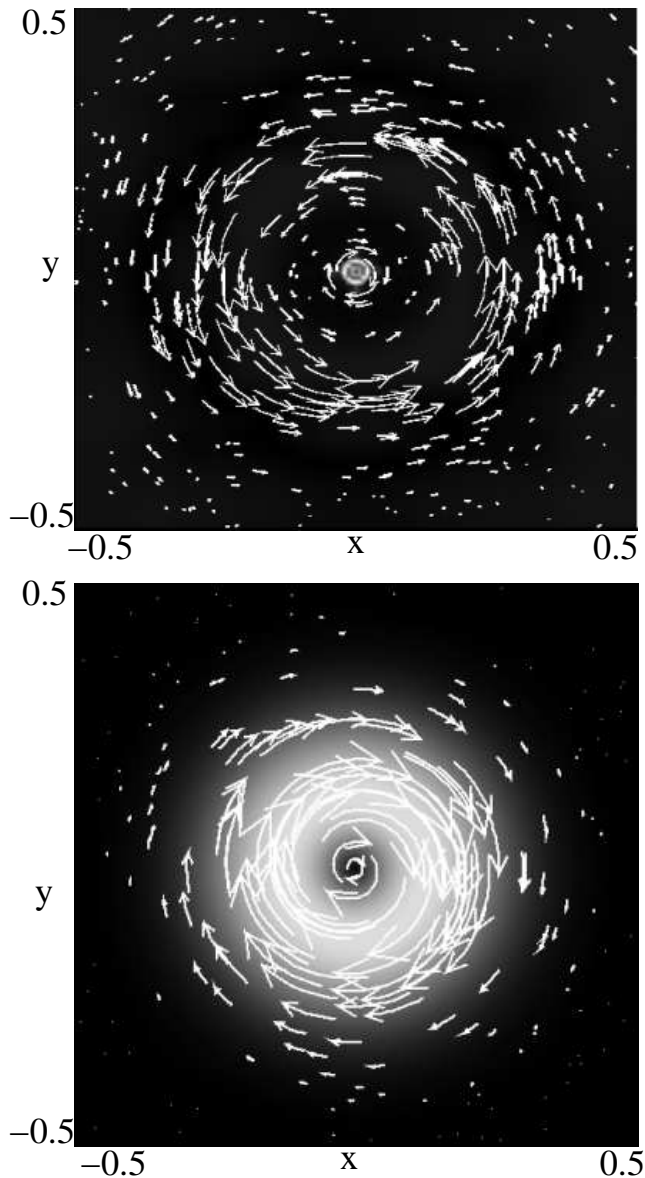}
\caption{Plasma flow in the plane $z \approx 0.05$, showing
  rotation centred on the spine (centre), for perturbations (a) rotation in the
  fan and (b)  rotation about the spine. 
  Pattern in other planes $z=z_0$ is similar, with
  only the $r$-localisation varying due to the hyperbolic nature of the flux
  enevelope affected by the perturbation. The background shading shows the current
  modulus in the same plane in each case. Each image is at the time of maximum
  current, and the parameters are as in Figure
  \ref{j_fanrot} for (a), and for (b) $B_0=1,
  b_0=\pm0.1,  \zeta_0=\pm 0.2, \beta=0.01, a_h=0.06, b_h=0.06$, $\eta=\eta_h$.}
\label{rotflow}
\end{figure}

An examination of the plasma flow induced by the perturbation is of interest
in itself, as well as in helping to determine what type of reconnective
behaviour might result once a current sheet forms. It is perhaps not
surprising that the rotational perturbation induces rotational plasma flows
centred on the spine,
in the $xy$-plane, as shown in Figure \ref{rotflow}(a). The rotation is
present  within the `envelope' of magnetic flux
which was magnetically connected to the initial perturbation.
Importantly, there is no plasma flow across either the spine or fan of the
null. 
In fact, the direction of the plasma rotation in Figure \ref{rotflow}(a)
is somewhat complicated by reflections
from the $z$-boundaries. Initially, the ingoing pulse generates
clockwise flow (viewed from the $z$-boundaries) and the outgoing pulse
anti-clockwise. In addition, once the ingoing pulse is reflected off the 
$z$-boundaries, it
generates an additional region of anti-clockwise flow, which propagates in
the negative $z$-direction. This occurs first at large radius within the flux
envelope since this is where the reflection first occurs
(see Figure \ref{j_fanrot}). Thus, present in Figure \ref{rotflow}(a) are the
flow resulting from the ingoing pulse, and its reflection from the $z$-boundary.

\begin{figure*}
\centering
\noindent\includegraphics[width=39pc]{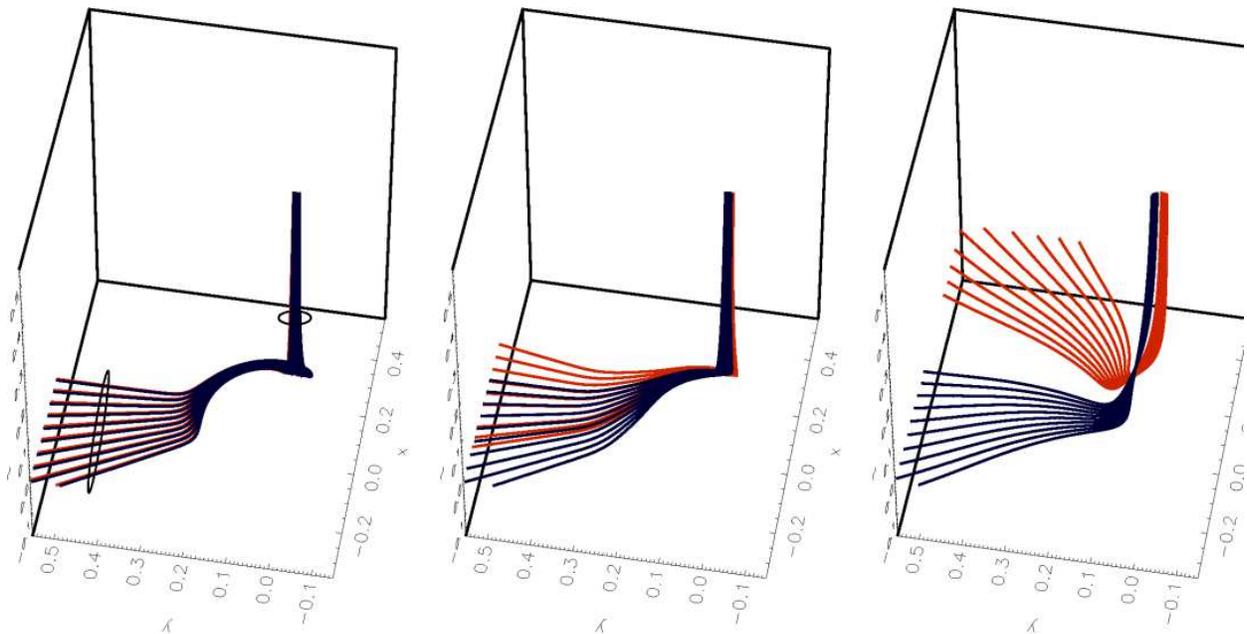}
\caption{Rotational slippage between two sets of magnetic field lines. The
  black loops enclose the two sets of fluid elements from which the field
  lines are traced. For parameters $B_0=1, b_0=3, R_{10}=0.18, \beta=0.01,
  a_h=0.06, b_h=0.06$ and constant $\eta=0.002$.
  Again the total domain size is $1.5\times1.5\times2.5$. Times for the plots
  are $t=0, 0.2, 0.9$.}
\label{frot_blines}
\end{figure*}

As a result of this plasma flow, it seems unlikely that any reconnection
involves field lines being advected across the separatrices, but rather a
rotational mismatching would be expected. This is predicted 
by kinematic studies \citep{pontinhornig2005} when the current is parallel
to the spine of the null point.  As shown in Figure \ref{frot_blines}, this is
indeed the case. The plotted field lines are traced from two sets of 
trace-particles, which are initially connected (see Figure
\ref{frot_blines}(a)). The first set of particles (from which the grey (blue online) field
lines are traced) remain approximately in the ideal region, i.e.~they are
located far away from the null (near the fan plane) where currents remain weak. 
These field lines therefore show approximately the ideal behaviour,
which is followed everywhere except close to the spine. (When
  the field lines pass 
close to the spine in the figure it is not possible to see the individual
lines, due to the converging field structure.)
The other set of particles
is located close to the spine, such that they are eventually engulfed by the
localising current. It can be seen that the corresponding field lines
(black, red online) continually change their connections
\citep[see][]{priesthornig2003} in a rotational fashion as the current
localises. Note that in the simulation pictured, a higher value of $b_0$, and
a constant 
resistivity ($\eta=0.002$, implying a Lundquist number of order 500) 
have been used for illustrative purposes, 
since the diffusive region is larger for
constant $\eta$ than with the hyper-resistivity, and the effect of the diffusion is
therefore easier to visualise.

\subsection{Parallel electric field}

\begin{figure}
\centering
\noindent\includegraphics[width=20pc]{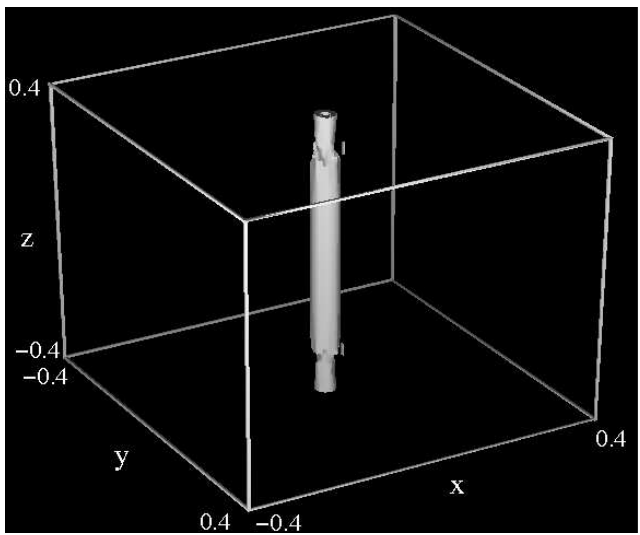}
\caption{Isosurface of $E_{\parallel}$ resulting from a rotation in the
    fan plane, at $50\%$ of its maximum, at $t=2.16$ and
  for the same parameters as Figure \ref{j_fanrot}.}
\label{epar_frot}
\end{figure}

It is well-known that a crucial indicator of 3D magnetic reconnection is 
the presence of an electric field component parallel to the magnetic field,
$E_{\|}$ \citep[e.g.][]{schindler1988}. Examining an isosurface of $E_{\|}$,
it is clear that non-ideal effects become important basically uniformly all
along the spine (see Figure \ref{epar_frot}; taken a little before current
maximum to limit appearance of boundary effects). Thus the reconnection seems
to be associated with a spatially diffuse (at least in one dimension) region.
 
\cite{pontin2004} showed that in the case of an isolated 3D null with current
parallel to the spine, the rate of rotational flux mis-matching can
be quantified by calculating the integrated parallel electric field (from one
end of the diffusion region to the other) along the spine of the
null (see also \cite{hornig2003}). 
Care must be taken, however, in comparing our simulations with this kinematic
result. The principal reason for this is that the result relies upon the
assumption that the diffusion region is bounded in $z$, encompassing the null.
Nonetheless, we find that $\Phi_s=\int E_{\parallel}$ (along the
whole length  of the spine within the domain) does indeed
show a clear peak in time, indicating a maximum in the reconnection rate
  at the null,
which occurs once the perturbation reaches the spine.
That is,
when the field gradients built up sufficiently that the current sheet begins
to diffuse onto the spine, a parallel electric field can be seen to develop
along it (see Figure \ref{j_fanrot}). 
While the numbers we obtain here (Figure \ref{j_fanrot}(b)) for $E_{\|}$ are
dependent on the resistivity model and/or value, the qualitative behaviour is not.

It is interesting to observe the close temporal correlation between the peak
reconnection rate ($\Phi_s$) at the null and the peak current. As mentioned
previously, in the ideal limit we would expect $J$ to grow indefinitely, due
to the structure of the magnetic field in the vicinity of the null. Thus a
current peak is also an indication (for our transient perturbations) that
non-ideal processes have become 
important, allowing the stress in the field (here twist) to begin to
significantly dissipate. 
With this type of perturbation, $E_{\parallel}$ will also increase as the 
pulse approaches the spine due to the geometry of 
the magnetic field, which becomes increasingly $B_z$-dominated as one approaches
the spine line (at a given height, $z$).

\section{Rotation about the spine}\label{spinerotsec}
For a rotation about the spine we may take a perturbation magnetic field of
similar form to that used in the previous section. However, here we take
$\zeta_0$ non-zero and $R_{10}$ to be zero. For perturbation above and below
the fan plane, we superpose two disturbances of this form, with $\zeta_0$ of
opposite signs (Figures \ref{perturb_pic}(c) and \ref{perturb_bl}(c)). Whether
the rotation is in  the same or opposite sense on either side of the fan is
determined by the relative signs of $b_0$. We take $B_0=1$, $b_0=\pm 0.1$,
$\beta=0.01$, $\zeta_0=\pm 0.2$, $a_h=b_h=0.06$.
A thorough examination of the development of these types of perturbations has
been carried out in a similar simulation, described by
\cite{galsgaardpriest2003}. 
The one major difference between the two simulations is that they applied
  a driving velocity at the (line-tied) boundaries, whereas we perturb the
  magnetic field within the domain.
We therefore only summarise the results here. 

\begin{figure}
\centering
\noindent\includegraphics[width=18pc]{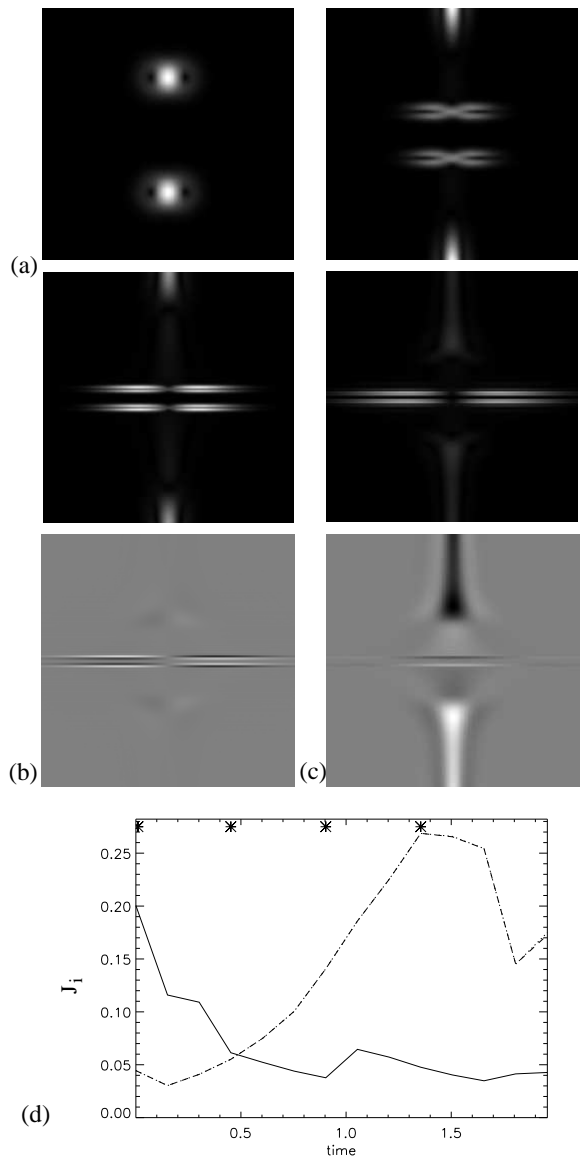}
\caption{For a rotation about the spine of opposite sense above and below
    the fan; (a) shaded images showing $|{\bf J}|$ in the $y=0$ plane, for
  $[x,z]=[\pm0.45,\pm0.45]$,  at times shown by the asterisks in (d) ($t=0,
  0.45, 0.9, 1.35$). (b) $J_x$ and (c) $J_z$ at $t=1.35$ (time of maximum
  current). (d) Evolution of the maximum values of each current component in
  the inner half (in each direction) of the domain: $J_x$ dotted, $J_y$
  dashed, $J_z$ solid line.  For domain size $1.5\times1.5\times 1.5$  and
  parameters as in Figure \ref{rotflow}(b).}
\label{j_spinerot}
\end{figure}

This time, as the perturbation evolves, it generates a current which
  spreads out in the  fan plane as it approaches the
  null (see Figure \ref{j_spinerot}(a)).
However, the most
important aspect of the current development is that there is once again no
tendency for a focusing of the localisation towards the null itself. Regardless of
whether the rotation has the same or opposite sense, there are inward and
outward propagating disturbances, and in the ingoing disturbance ${\bf J}_{xy}$ 
is amplified (Figure \ref{j_spinerot}(b,d)). 
The outgoing pulse, which localises towards the spine in the same way as
  was seen in Section \ref{fanrotsec}, shows a growth of $J_z$ (Figure
  \ref{j_spinerot}(c)).  However, this
  time that current never reaches the null since the fan is not perturbed.
There is also a relatively weak pulse of $J_z$ which does propagate
  towards the fan plane and null (\ref{j_spinerot}(c)), as seen by
  \cite{rickard1996}, however, this is 
  dominated by the strong ${\bf J}_{xy}$ further out in the fan.

The fact that different current components are magnified in the ingoing and
  outgoing pulses is likely just a result of
the geometry of the background field. In particular, moving inwards from a given point on
the spine, $d/dx, d/dy$ decrease, while $d/dz$ increases sharply as the fan is
approached. Thus the
  current maximum occurs as the ingoing pulse steepens towards the fan. If the
  resistivity were zero, a fan current sheet (with infinite current density but of
  zero thickness) would be asymptotically approached. However,
  in our simulation, sufficiently steep gradients develop before the
  disturbances reach the fan, such that diffusion becomes important. This is
  one difference between 
  our simulation and the driven case of \cite{galsgaardpriest2003}, where the
  continual twisting at the boundaries drove the current right into the fan plane.

The plasma flow in this case is, as expected, of a rotational nature, with
similar form to that found for rotation in the fan plane (and sense of
rotation governed by the initial condition). A plot of the plasma velocity in
the $xy-$plane demonstrating this is shown in Figure \ref{rotflow}(b).
Furthermore, the field line behaviour is found to be very similar to that
shown in Figure \ref{frot_blines}.

As the disturbances here travel basically along ${\bf B}$ and there
is no flow and no magnetic connection between the two sides of the fan plane
($z>0, z<0$), there is essentially no difference for the case where the
driving has the same sign on each side of the fan, except the sense of
rotation. In addition, since
the current is not driven right into the fan as in \cite{galsgaardpriest2003},
there is no issue of the currents reinforcing or cancelling there as they found.

\section{Shear of the fan plane}\label{fanshsec}
The behaviour of shearing-type perturbations is very different to rotational
ones. We first consider the case of a shear of the fan plane. The perturbation
again takes the form described by Equation (\ref{perturb}), where this time
$\{b_i, b_j\}=\{b_x, b_z\}$ and $\zeta=y$. In addition we take
$R_1=\sqrt{(x-x_0)^2 +z^2}$, $\theta_1=tan^{-1}\left( \frac{z}{x-x_0} \right)$
and $R_{10} = \zeta_0 =0$. 
This corresponds to a linear tube (of finite length) of azimuthal magnetic
flux, whose axis lies along $x=x_0, z=0$.
We superimpose two such perturbations on our
background null, with opposite signs of $x_0$, as shown in Figure
\ref{perturb_pic}(b). The effect on the fan field
lines is as if they have been `plucked', in one direction in some region of
the fan, and in the opposite direction in an opposite region
(Figure \ref{perturb_bl}(b)). We take (except where stated) $B_0=1$, $b_0=0.1$,
$\beta=0.01$, $x_0=\pm 0.16$, $a_h=0.06$, $b_h=0.2$.

\subsection{Current and plasma flow}
\begin{figure}
\centering
\noindent\includegraphics[width=20pc]{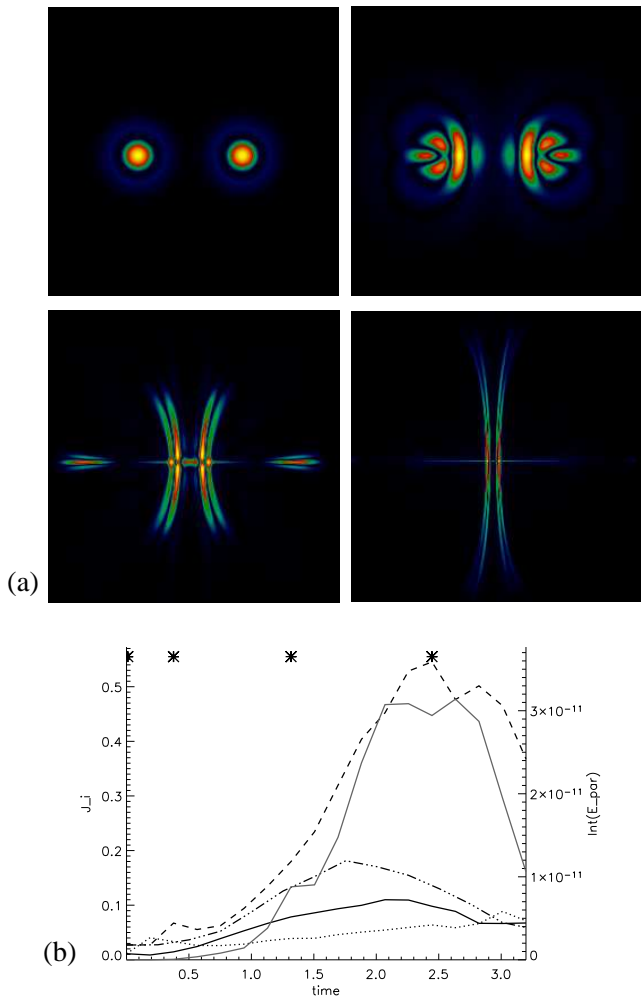}
\caption{
For a shear of the fan; (a) shaded images showing $|{\bf J}|$ 
in the $y=0$ plane, for $[x,z]=[\pm0.45,\pm0.45]$, at times marked by
  asterisks in (b) ($t=0, 0.38, 1.32, 2.45$). (b) Evolution of the maximum
  values of each current component in the inner half (in each direction) of
  the domain: $J_x$ dotted, $J_y$ dashed, $J_z$ solid black line. The grey
  line plots the integrated parallel electric field along the $y$-axis.  For
  domain size $1.5\times1.5\times1.5$ and parameters $B_0=1,  b_0=0.1,
  x_0=\pm 0.16, \beta=0.01, a_h=0.06, b_h=0.2$,
  $\eta=\eta_h$. The dot-dashed line shows $J_y$ evolution in
  a run with the same parameters, but  with $\eta=5\times 10^{-5}$, constant}.
\label{j_fsh}
\end{figure}

The crucial characteristic of the current development for a shear of the fan
is that, while the disturbance propagates to some extent along the
background field lines, there is additionally a strong focusing of the
current (across the field) towards the null point itself (see Figure \ref{j_fsh}(a)).
Note that again there is a fair degree of structure to the localising
  current. This is because the disturbance was initiated by a
  divergence-free magnetic disturbance, in the form of a tube of magnetic
  field (see Figure \ref{perturb_pic}(b)). Thus in fact three current
  concentrations, demarking field gradients at the front, middle and back end
  of the pulse are present. The components at the front and back end are
  necessarily of opposite sign (in $J_y$) to the strongest concentration in
  the middle, and the gradients of course go through zero between these
  regions, hence the appearance of three `pulses' in $|{\bf J}|$. While the
  two outer regions initially join up to encircle the inner concentration, this
  appearance is eventually lost due to the stretching of the structure in $z$.

It is evident from Figure \ref{j_fsh}(b) that the disturbance which focuses
towards the null point is dominated by $J_y$, that is, the current orthogonal
to the plane of the shear. 
Once again it is the resistivity which limits the growth of the
  current. In an ideal situation we would expect the current to continue to
  grow indefinitely, focusing at the null, where a good portion of the energy of the
  disturbance would be deposited, once resistive effects eventually become important.

As an explicit demonstration of the effect of our hyper-resistivity
  model, we re-run the simulation with $\eta$ constant, taking $\eta=5\times 10^{-5}$
  (in order to resolve all structures sufficiently). We find that the
  diffusion causes the peak current to be much lower, and occur much earlier when the
  perturbation is far less localised (see the dot-dashed line in Figure
  \ref{j_fsh}(b)). Qualitatively the current structure 
  behaves very similarly---the same images as in Figure \ref{j_fsh}(a) look
  almost identical (when scaled by their individual maxima), except that in
  the final image (now long after current maximum), the current is not so
  strongly peaked near $z=0$ as much of the perturbation has diffused away.


\begin{figure}
\centering
\noindent\includegraphics[width=20pc]{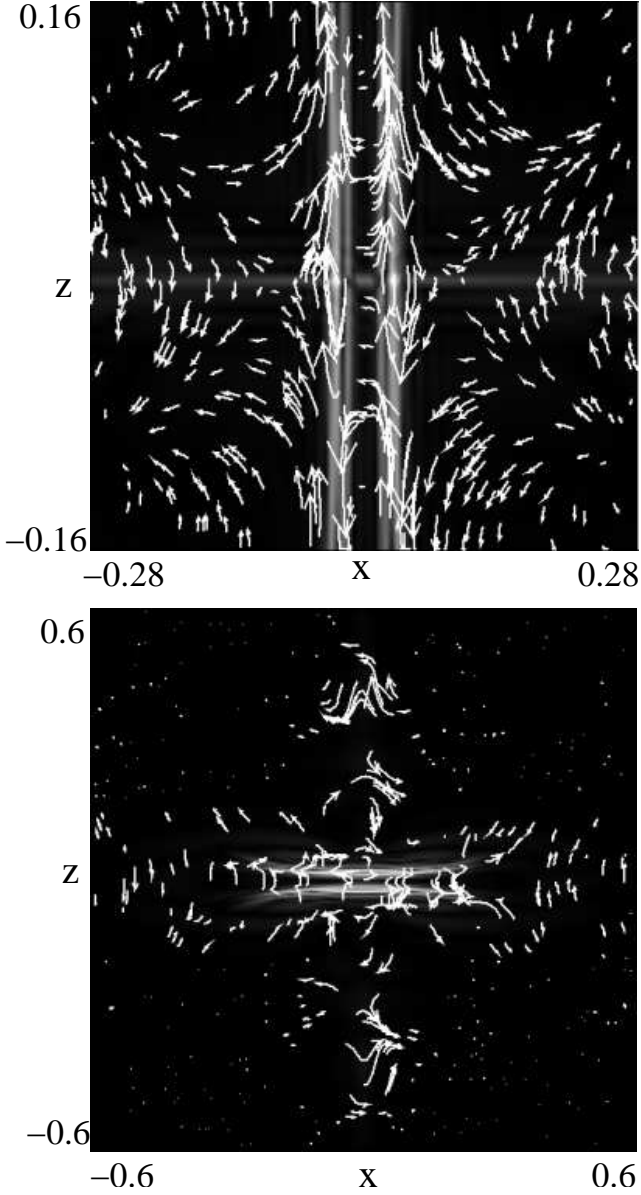}
\caption{(a) The plane $y=0$, perpendicular to the shear, at time of maximum
  current ($t=2.45$), for a shear  of the fan. The arrows show the
  plasma flow, and the background shading the current modulus. 
  Parameters as in Figure \ref{j_fsh}. (b) As (a), but
  for spine shear run ($t=1.23$), for domain size $1.5\times1.5\times1.5$
  and parameters $B_0=1, b_0=0.1, z_0=\pm 0.2, \beta=0.01, a_h=0.05, b_h=0.05$, $\eta=\eta_h$.}.
\label{v_sh}
\end{figure}

The presence of strong current at the null point directed
parallel to the fan is expected to indicate the presence of more
  traditional reconnection-type flows. In the simplified kinematic
model of \cite{pontinhornig2005}, a stagnation-point flow, centred on the
null, is present in the plane perpendicular to the shear. This flow transports
magnetic flux across the spine and fan. Similarly here, 
at the time of maximum current (see
Figure \ref{v_sh}(a)), we indeed find strong plasma flow across the fan plane
(originally $z=0$), concentrated around the areas of maximum current. 
Away
from the current concentrations as well, there is flow across the original
locations of the spine and fan.  
However, it is not  straightforward to determine whether
magnetic reconnection is occurring at the null. This is because the flow may act to
transport magnetic flux across the fan plane (magnetic reconnection), or it
may simply act to advect the spine and fan in an ideal sense. 
It is therefore again of interest to examine the development of
$E_{\parallel}$.

\subsection{$E_{\parallel}$ and reconnective behaviour}
The presence of a parallel electric field denotes a breakdown of ideal
  behaviour, and so is crucial in determining whether magnetic reconnection occurs.
\begin{figure}
\centering
\noindent\includegraphics[width=20pc]{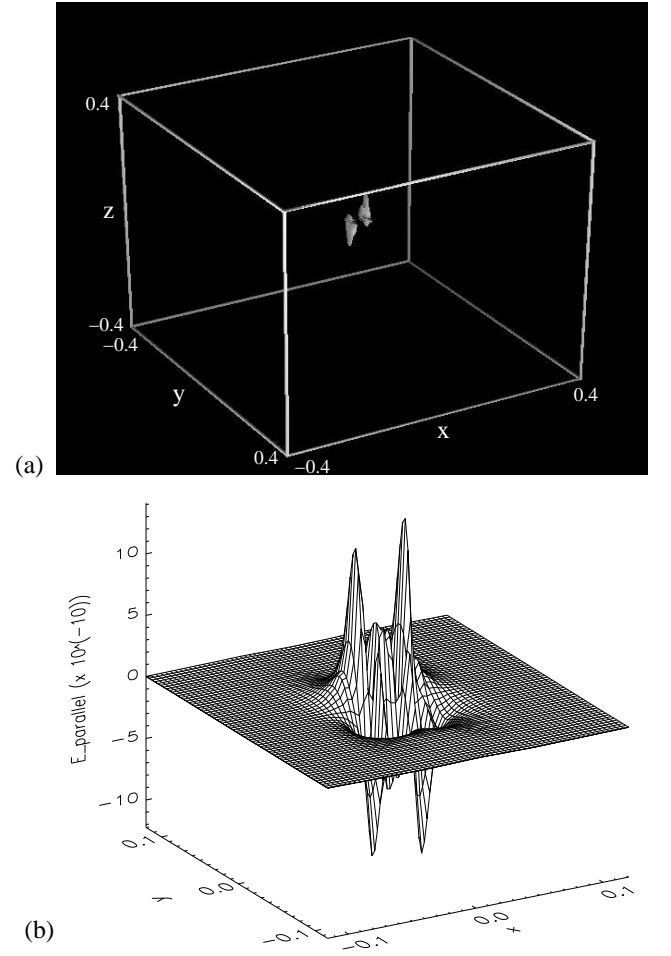}
\caption{For a shear of the fan plane, (a) isosurface of $E_{\|}$ at $25\%$ of its
  maximum and (b) surface of $E_{\|}$ in the $z=0$ plane (interpolated
  onto a uniform grid), both at  $t=2.45$ (current
  maximum), for the same parameters as in Figure \ref{j_fsh}.}
\label{epar_fsh}
\end{figure}
We expect $E_{\parallel}$ to be greatest in regions of high current, where
the field lines lie in the direction of the current (since $E_{\parallel}=\eta
J_{\parallel}$). Examining an isosurface of $E_{\|}$ at current maximum
(Figure \ref{epar_fsh}), we
see that in fact the highest concentrations are located (as expected) at
regions of strong $J$, a little way from the $y$-axis (due to the structure of
the initial disturbance). The most striking thing we see from Figure \ref{epar_fsh} is
that in this case the regions in which $E_{\|}$ develops are highly spatially
localised in all directions. Hence reconnection processes are very local to
the null itself. While this localisation is not quite so extreme in runs with
constant resistivity, the basic structure is the same.

For an isolated null with fan-aligned current, surrounded by a localised non-ideal
region, the
integral of $E_{\|}$ along the magnetic field line in the fan plane which is
directed along ${\bf J}$ can be shown to give an exact measure of the rate of
flux transfer across the fan, where the integral is taken from one side of the
diffusion region (assumed spatially localised at the null) to the other
\citep{pontinhornig2005}.  
Thus, a non-zero value for $\Phi_f=\int E_{\|}ds$ along a fan field line
  threading the diffusion region implies that magnetic flux is transferred
  across the separatrix (fan) plane, even in the ideal region, and thus
  demonstrates that reconnection is occurring at the null.
Performing
such an integration, $\Phi_f=\int_{x=z=0}E_{\parallel}ds$, we find a parallel
electric field does indeed develop, a strong
indicator that some reconnective process is taking place. 
Note that here again
the peaks of $\Phi_f$ and $J_{max}$ are closely correlated.

In fact, when $\Phi_f$ reaches its temporal maximum, the spatial maximum of
$E_{\parallel}$ is still some distance from the $y$-axis (see Figure
\ref{epar_fsh}). In the fan plane, the maxima actually occur on field lines
which lie approximately along $x=\pm y$.  
The reason for this is as follows. The electric field, like ${\bf J}$, is
  directed largely in the ${\hat {\bf y}}$ direction perpendicular to the
  shear, and is uni-directional
  through the null. Thus $E_{\|}\approx 0$ on the $x$-axis, and $E_{\|}>0$
for $y>0$, $E_{\|}<0$ for $y<0$ (see Figure \ref{epar_fsh}(b)). 
If the perturbation were to extend
azimuthally around the entire fan plane, we would expect the maximum and minimum of
$E_{\|}$ to be on the $y$-axis. 
Such an azimuthally symmetric perturbation could be envisaged by
  considering a field localised within the torus of Figure
  \ref{perturb_pic}(a) which is purely poloidal rather than toroidal, and
  which has a $cos(\theta)$ multiplying factor (where $\theta=tan^{-1}(y/x)$).
However, since the chosen shear has a finite azimuthal
extent, maxima and minima of $E_{\|}$ are found off the $y$-axis. 
Therefore, by comparison with 
\cite{pontinhornig2005}, we expect a better measure of the rate of flux
transfer across the fan to be given by ${\Phi_f}^{\prime}=
\int_{x=y,z=z_{fan}}E_{\parallel}ds$, by symmetry, giving the maximal value of
$\Phi_f$. This quantity shows a similar qualitative evolution to that of $\Phi_f$ plotted
in Figure \ref{j_fsh}(b).

\begin{figure*}
\centering
\noindent\includegraphics[width=39pc]{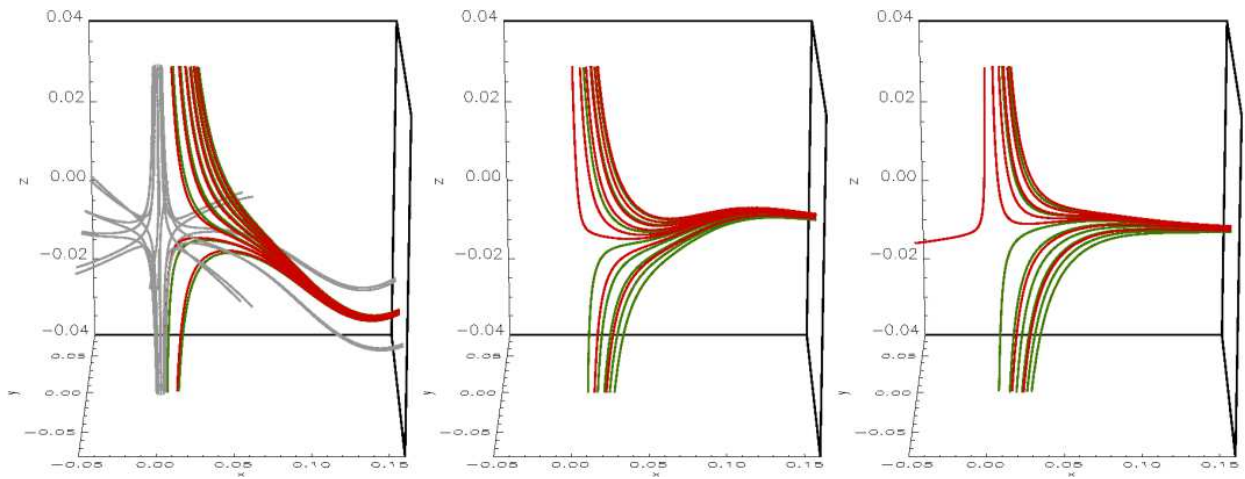}
\caption{Evolution of two initially connected sets of field lines when the fan
  is sheared. One
  set are traced from plasma elements which remain always in the ideal region
  far up the spine (black, red online) and is advected across the spine. The other
  (grey, green online, traced from plasma elements initially within the black loop) are
  traced from plasma elements  near the $z=0$ plane and are
  transported  across the fan. Parameters as
  in Figure \ref{j_fsh}, but $b_0=3, \eta$
  constant $=0.002$. The dashed field lines in the first image indicate the
  initial location of the spine and fan. Times for the images are $t=0, 0.8, 1.6$}
  \label{fsh_blines}
\end{figure*}

The nature of the magnetic reconnection associated with $E_{\|}$ can again be
examined by tracing field lines from trace particles which are initially
magnetically connected. A typical evolution for an appropriately chosen set of
field lines is shown in Figure \ref{fsh_blines}. It is clear that the chosen
plasma elements change their connections in such a way as to suggest a
transport of magnetic flux across both the spine and fan of the null. 
Specifically, grey field lines (green online, traced from `non-ideal' footpoints
located near the fan plane, between the initial perturbation and the
null) are transported  across the fan plane (from above to below), whereas black field
  lines (red online, traced from `ideal' plasma elements located far up the spine)
  are advected across  the spine.

\section{Shearing the spine}\label{spineshsec}
To complete the picture we finally consider a perturbation which shears the
spine of the null point. The disturbance takes the form given by Equation
(\ref{perturb}), with $\{b_i, b_j\}=\{b_x, b_z\}$, $\zeta=y$,
$R_1=\sqrt{x^2 +(z-z_0)^2}$, $\theta_1=tan^{-1}\left( \frac{z-z_0}{x} \right)$ 
and $R_{10} = \zeta_0 =0$. Again two such disturbance fields are superimposed,
with opposite signs of $z_0$ (see Figure \ref{perturb_pic}(d)). The effect on
the spine field line is as if it has been locally `plucked', in opposite
directions above and below the null (see Figure \ref{perturb_bl}(d)). We
  take (except where stated) $B_0=1$, $b_0=0.1$, 
$\beta=0.01$, $z_0=\pm 0.2$, $a_h=b_h=0.05$.


\begin{figure}
\centering
\noindent\includegraphics[width=20pc]{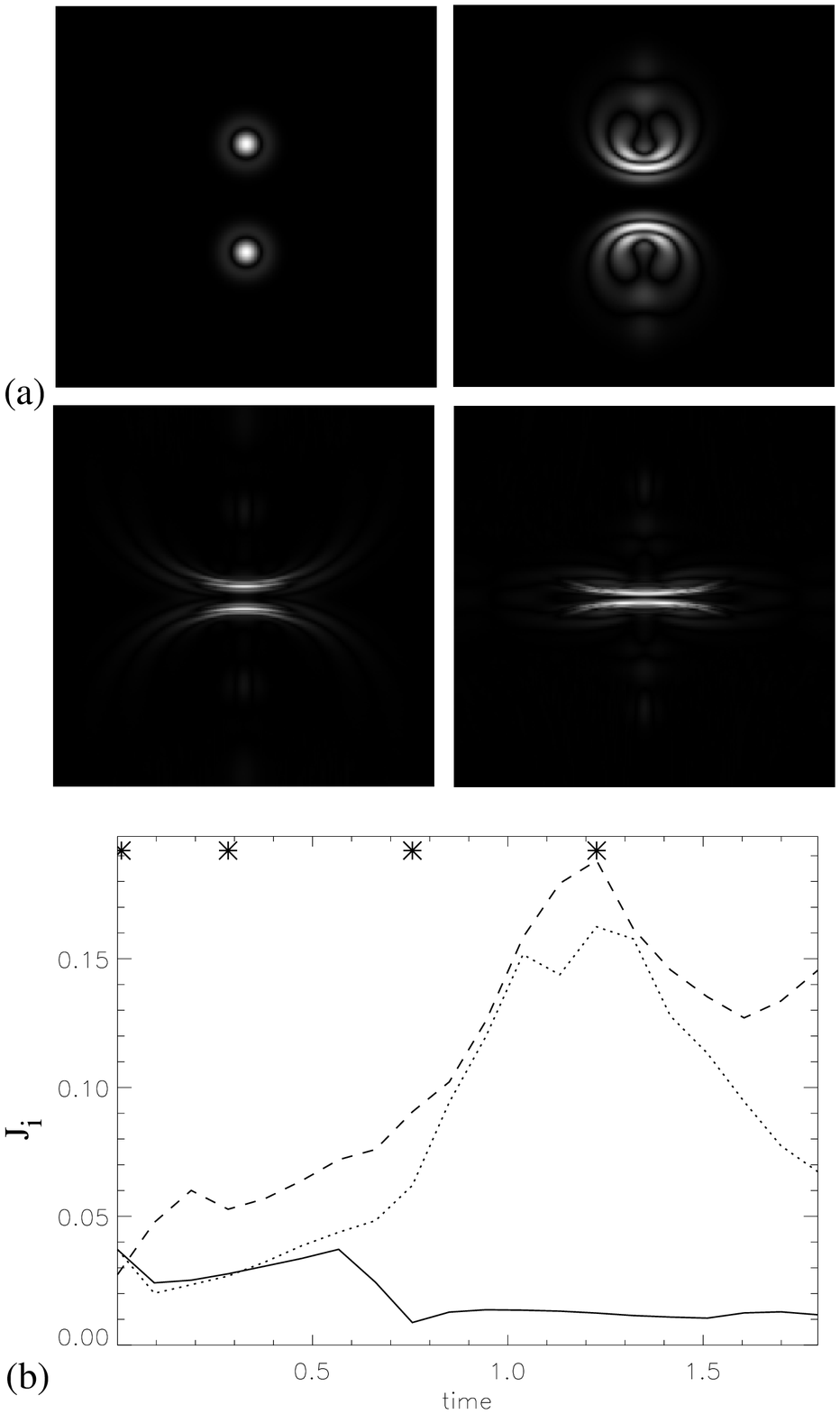}
\caption{For a shear of the spine; (a) shaded images showing $|{\bf J}|$
  in the $y=0$ plane,  for
  $[x,z]=[\pm0.75,\pm0.75]$, at times marked by asterisks in (b)
  ($t=0, 0.28, 0.76, 1.23$). (b) Evolution of the maximum values of each current component in
  the inner half (in each direction) of the domain: $J_x$ dotted, $J_y$
  dashed, $J_z$ solid black line. Parameters as in Figure \ref{v_sh}(b).}
\label{j_spsh}
\end{figure}

The resulting
current evolution in the plane perpendicular to the shear is shown in Figure
\ref{j_spsh}, from which it can be seen that the inward-propagating component
of the disturbance is again attracted towards the null point, as in the case
of a shear of the fan. Although both
$J_x$ and $J_y$ are magnified during this localisation, in fact ${\bf J}=J_y$
at the null point itself. That is, the current which develops at the null is
again parallel to the fan, perpendicular to the plane of shear.

As before, a plot of the plasma velocity in the plane perpendicular to the
shear (see Figure \ref{v_sh}(b)), suggests flow across the spine and fan,
although again it is hard to say whether the flow actually crosses the spine
and fan, or merely ideally advects them.
In addition, there is also some complicated (and relatively strong) flow
across the spine within the region of the localising current concentration.

\begin{figure*}
\centering
\noindent\includegraphics[width=39pc]{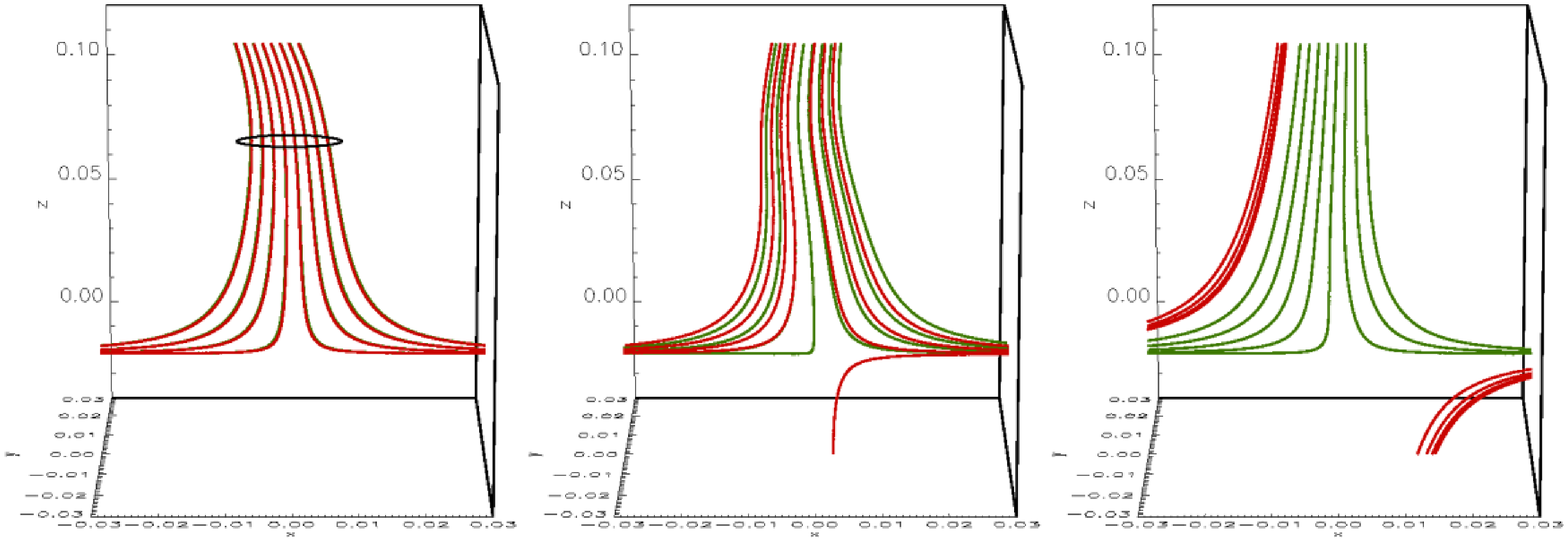}
\caption{For a shear of the spine, evolution of field lines traced from
  plasma elements near the spine (black, green online, `non-ideal', traced from plasma
  elements initially located within the black ring in the first image) and
  near the fan (grey, red online, `ideal'). Parameters as in Figure \ref{v_sh}(b), except
  that $b_0=3$, $\eta=0.002$ constant. Images are at times $t=0, 0.25, 0.7$.} 
\label{spsh_blines}
\end{figure*}

Some representative field lines are plotted in Figure \ref{spsh_blines}.
Field lines traced from plasma elements initially located near the spine
(black, green online) are transported back and forth across the spine, since these plasma
elements are engulfed by the current concentration. The grey field lines (red online),
however, are traced from plasma elements which stay forever in the `ideal'
region, far out along the fan plane, and are clearly
advected across the separatrix surface. In fact, the behaviour of the null in
response to the perturbation is similar to the case where the fan plane is
sheared. In each case, the current is peaked at the null itself, and 
field lines reconnect in a manner
similar to the traditional 2D picture, in the plane of the shear.

\section{Summary}\label{summary}
The behaviour of perturbations in the vicinity of a potential 3D magnetic null
point has been investigated via resistive magnetohydrodynamic
simulations. Disturbances which affect the null point
 must deform the field in the vicinity of either the spine or the fan of the
 null.  
A typical such disturbance may be constructed using a combination of
shears and rotations, and hence we considered four basic possibilities---namely
shear of the fan or spine, and rotation of the fan or about the spine.
It should be pointed out that we consider a case with no initial
  `background' flow, that is there is no flow through our
  boundaries, and therefore we do not compare with strongly driven flux pile-up
  models.

We have found that rotational types of perturbations tend to lead to current
accumulation in the vicinity of the field lines which asymptotically approach
(or recede from) the null point, that is the spine and fan field lines. The
disturbances behave in a way which is essentially Alfv{\' e}nic, propagating
along the background field lines. Thus, (apart from a weak non-linear
wave-mode coupling) there is no  preferential growth of current at the null
point itself.  Near the null, the current accumulates along the spine when the
rotations  disturb the fan plane, while rotations about the spine lead to
currents in the location of the fan near the null.  Such diffuse currents,
extending along the separatrix surfaces, have been predicted in incompressible
analytic models \citep{craigetal1995,craig1996}, although the form we find for
the current sheets is somewhat different. The plasma flows and field line
reconnection associated with these currents is of a rotational nature,
as predicted by \cite{pontin2004}.

By contrast, when the spine or fan of the null is subjected to a shearing
perturbation, there is a strong current growth which is localised at the null
point itself. This is achieved by propagation of the disturbance across the
background magnetic field lines, suggesting that the dominant wave mode
associated with the disturbance is a fast mode. This behaviour agrees with the
results of \cite{rickard1996}, and is  consistent also with the work of
\cite{pontincraig2005}, who found singular current sheets to result at a 3D null point
in an ideal line-tied relaxation when a shear was applied to spine or fan.
The current which develops at the null is dominated by the component in the
direction perpendicular to the plane of shear.
The resulting plasma flow has a stagnation-point structure. It is in general
  very difficult to say from such simulations definitively that the plasma
  actually crosses the spine and fan. In the 
ideal region the flow advects field lines ideally, however it may still cross
the spine or fan, if the spine and fan field lines change their identities
(where now a field line is identified by plasma 
elements which lie on it in the ideal region) in
time via magnetic reconnection (as in steady-state 2D reconnection). Despite
this difficulty, by 
comparison with kinematic models \citep{pontinhornig2005}, the non-zero
integrated parallel electric field which develops along fan field lines is a
strong indicator of flux transfer across at least the fan.
The magnetic field line evolution
shows a reconnection of field lines through/around the spine and across
the fan plane, as in the `spine reconnection' and `fan reconnection' of
\cite{priest1996}. The transport of magnetic flux across the fan plane is a
particularly important property, as it is a separatrix surface of the magnetic
field. The behaviour of shearing perturbations is further in agreement with
the kinematic predictions of \cite{pontinhornig2005}.

Finally, we investigated the existence and behaviour of parallel electric
  fields within the simulations.  In each case investigated, in order to halt
  the indefinite growth of ${\bf J}$, resistive effects become important, as
  evidenced by the development of $E_{\|}$. In the case of rotational
  perturbations, the spatial profile of $E_{\|}$ (as with ${\bf J}$), is
  spread out either along the spine or the fan of the null. For shear
  perturbations, however, $E_{\|}$ tends to be strongly localised at the null
  point itself.  Such parallel electric fields can be shown to give a
  physically meaningful measure of the reconnection rate at a 3D null
  \citep{pontin2004,pontinhornig2005} (though direct analogy with these
  kinematic results is not completely straightforward here). Finally, the
  dependence of the behaviour of the system, and the scaling of the peak
  currents and reconnection rate, on resistivity, plasma-$\beta$ and the
  isotropy of the background null point are all important issues, as is
  the energetics of the reconnection process. These will
  be left for exploration in a future paper.

\begin{acknowledgements}
The authors would like to acknowledge helpful discussions with
A.~Bhattacharjee and I.~J.~D.~Craig. D.~P.~was funded by 
the Department of Energy, Grant No.~DE-FG02-05ER54832, by the
National Science Foundation, Grant No.~ATM-0422764, 
and in the initial stages of this work by the Marsden Fund,
grant no.~02-UOW-050 MIS. K.~G.~was supported by the Carlsberg Foundation in
the form of 
a fellowship. Computations were performed on the Zaphod Beowulf
cluster which was in part funded by
the Major Research Instrumentation program of the National Science Foundation under
grant ATM-0420905.
\end{acknowledgements}



\end{document}